# Non-Linear Realization of $\aleph_0$-Extended Supersymmetry[1]

Hitoshi NISHINO[2]

*Department of Physics*
*University of Maryland at College Park*
*College Park, MD 20742-4111, USA*

**Abstract**

As generalizations of the original Volkov-Akulov action in four-dimensions, actions are found for all space-time dimensions $D$ invariant under $N$ non-linear realized global supersymmetries. We also give other such actions invariant under the global non-linear supersymmetry. As an interesting consequence, we find a non-linear supersymmetric Born-Infeld action for a non-Abelian gauge group for arbitrary $D$ and $N$, which coincides with the linearly supersymmetric Born-Infeld action in $D = 10$ at the lowest order. For the gauge group $U(\mathcal{N})$ for M(atrix)-theory, this model has $\mathcal{N}^2$-extended non-linear supersymmetries, so that its large $\mathcal{N}$ limit corresponds to the infinitely many ($\aleph_0$) supersymmetries. We also perform a duality transformation from $F_{\mu\nu}$ into its Hodge dual $N_{\mu_1\cdots\mu_{D-2}}$. We next point out that any Chern-Simons action for any (super)groups has the non-linear supersymmetry as a hidden symmetry. Subsequently, we present a superspace formulation for the component results. We further find that as long as superspace supergravity is consistent, this generalized Volkov-Akulov action can further accommodate such curved superspace backgrounds with local supersymmetry, as a super $p$-brane action with fermionic kappa-symmetry. We further elaborate these results to what we call 'simplified' (Supersymmetry)$^2$-models, with both linear and non-linear representations of supersymmetries in superspace at the same time. Our result gives a proof that there is no restriction on $D$ or $N$ for global non-linear supersymmetry. We also see that the non-linear realization of supersymmetry in 'curved' space-time can be interpreted as 'non-perturbative' effect starting with the 'flat' space-time.

---

[1]This work is supported in part by NSF grant # PHY-93-41926.
[2]E-Mail: nishino@nscpmail.physics.umd.edu

# 1. Introduction

The study of non-linear supersymmetry realizations has a very long history dating to the first effort by Volkov and Akulov [1], who gave the initial example of supersymmetry in four-dimensions ($D = 4$). This initial work was further elaborated by its linearization [2], and generalized to extended local supersymmetries in $D = 4$ [3]. The importance of non-linear realizations of supersymmetry has been revealed in many different contexts. For example, during the first revolution of string theory [4], there was strong motivation for supersymmetrization [5] of the Born-Infeld action [6][7], and initial results were summarized in [8] where the lower-order terms were explicitly and systematically presented.

The recent discovery of gauge-fixed supersymmetric Born-Infeld action in $D \leq 10$ to all orders [9] has drawn much attention in the context of D-brane dynamics [10][11][12][13][14] related to superstrings [4] and M-theory [15]. The non-Abelian generalization of these formulations was also considered in [16]. In such a formulation, there is already non-linear realization of supersymmetry, interpreted as the signal of 'broken' supersymmetry. In $D = 4$ formulations, non-linear supersymmetry has been interpreted in terms of partially broken $N = 2$ supersymmetry [17], with Nambu-Goldstone (NG) multiplets. The relationship between the non-linear and linear supersymmetry in superspace has been also studied in [18]. These global supersymmetric systems have been further applied to the question of spontaneous breaking of local supersymmetry (supergravity) [19].

Despite of these developments over the decades since 1970's, however, including the recent D-brane models, there still has been no universal guiding principle that can systematically yield new invariant actions for non-linear realization of supersymmetry in arbitrary space-time dimensions. This seems true also for the interpretation of non-linear realization as 'broken' supersymmetry [17]. On top of this, even though the original spirit of Volkov-Akulov formulation of non-linear supersymmetry [1] was 'off-shell' without auxiliary fields present and no physical superpartner fields in $D = 4$, this aspect has not been maintained in many formulations recently presented. Perhaps one of the reasons is that D-brane theory [10][13][9] necessitates the superpartner vector field for the gaugino.

In the case of linear realizations of supersymmetry, it is well-known that there are restrictions on the space-time dimensions $D$ and the number $N$ of extended supersymmetries, such as $N \leq 8$ in $D = 4$. However, very little has been stated about such restrictions on $D$ or $N$ for extended supersymmetries in the case of non-linear realizations.

In this paper, we will take a step toward a universal understanding of non-linear realization of supersymmetry in arbitrary dimensions. We will present simple actions that are invariant under non-linearly realized supersymmetry. We also give some matter lagrangians that are invariant under such global but non-linear supersymmetry. Such matter lagrangians coupled to NG-fermion within $D = 4$ have been studied in many contexts of phenomenological models [20], or non-linear realizations for extended supersymmetries [3]. We stress that our result is valid in any space-time dimensions $D$ as well as any number $N$ of extended supersymmetries. We find results applicable to arbitrary non-linearly realized $N$-extended supersymmetries, as long as the space-time allows the definition of spinor fields. As a by-product, we also give a supersymmetric Born-Infeld action for an arbitrary non-Abelian gauge group $\forall D$ and $\forall N$. We also establish a superspace formulation for these component results, whose geometrical significance is clearer than in a component formulation. We further give the super $p$-brane [21] interpretation of some of these actions, on supergravity backgrounds. This will be achieved by introducing the maximal-rank superpotential $C_{A_1 \cdots A_D}$ in an arbitrary even dimensions $D = 2k$, which was first formulated in the context of super eight-brane [22]. Motivated by this, we also present a possible (Supersymmetry)$^2$-formulation [23], in which the non-linear and linear realizations of supersymmetries coexist simultaneously within the same superspace.

## 2. Non-Linear Realization of Supersymmetry

### 2.1 Generalized Volkov-Akulov Action

We first setup our supersymmetry transformation which is the starting point in this paper. As mentioned above, there is no restriction on the space-time dimensions $D$ in our formulation, except that we have the usual signature $(D-1,1)$ with the flat metric $(\eta_{mn}) = \text{diag.}(-,-,\cdots,-,+)$, with only one time coordinate. Our system initially has only one NG-fermion, which may be Majorana, Weyl, Majorana-Weyl, or Dirac spinor. As will be seen, there is no restriction on such spinorial property for our formulation.

The rule for a supersymmetry transformation of the NG-fermion is essentially the same as those in [1] or [9], but we have to keep in mind that the space-time dimension $D$ is arbitrary:[3]

$$\delta_Q \lambda = \epsilon + i(\overline{\epsilon}\gamma^\mu \lambda)\partial_\mu \lambda \equiv \epsilon + \xi^\mu \partial_\mu \lambda \ , \tag{2.1}$$

where the parameter $\epsilon^{\underline{\alpha}}$ of the supersymmetry generator $Q_{\underline{\alpha}}$ does not depend on space-time coordinates $x^\mu$, since we are dealing only with global supersymmetry. The quantity $\xi^\mu$ defined by[4][5]

$$\xi^\mu \equiv i(\overline{\epsilon}\gamma^\mu \lambda) = -i\epsilon^{\underline{\alpha}}(\gamma^\mu)_{\underline{\alpha}\underline{\beta}}\lambda^{\underline{\beta}} \ , \tag{2.2}$$

behaves like the parameter for general coordinate transformation. It is easy to see the closure of supersymmetry on $\lambda$, as a routine check for consistency of the system:

$$[\,\delta_Q(\epsilon_1), \delta_Q(\epsilon_2)\,]\lambda = 2i(\overline{\epsilon}_2 \gamma^\mu \epsilon_1)\partial_\mu \lambda = \delta_P(2i\overline{\epsilon}_2 \gamma^m \epsilon_1)\lambda \equiv \delta_P(\zeta_{12}^m)\lambda \ , \tag{2.3}$$

---

[3]We do not introduce any dimensionful constants in this paper. We just put dimensionful constants like the Planck constant to unity.

[4]Here the gamma-matrix $\gamma^\mu$ is the same as $\gamma^m \delta_m{}^\mu$. The usage of the index $\mu$ instead of $m$ becomes clearer shortly.

[5]Here the underlines for the indices $\underline{\alpha}, \underline{\beta}, \cdots$ are for any additional indices for spinors, such as 'dotted-ness' for chiralities, or $USp(N)$-indices needed in some space-time dimensions $D$ [24]. Accordingly, the $\gamma$-matrix may also contain any relevant metric for such indices.

where $\delta_P$ is the usual translation, and $\zeta_{12}^m \equiv 2i(\overline{\epsilon}_2 \gamma^m \epsilon_1)$, reflecting the underlying supersymmetry algebra[6]

$$\{Q_{\underline{\alpha}}, Q_{\underline{\beta}}\} = 2(\gamma^m)_{\underline{\alpha}\underline{\beta}} P_m ~~, \tag{2.4}$$

A fundamental observation is that the quantity defined by

$$E_\mu{}^m \equiv \delta_\mu{}^m + i(\overline{\lambda}\gamma^m \partial_\mu \lambda) \equiv \delta_\mu{}^m + \Lambda_\mu{}^m ~~, \tag{2.5}$$

transforms under $\delta_Q$ as

$$\begin{aligned}\delta_Q E_\mu{}^m &= i(\overline{\epsilon}\gamma^\nu \lambda)\partial_\nu E_\mu{}^m + i(\overline{\epsilon}\gamma^\nu \partial_\mu \lambda)E_\nu{}^m \\ &= \xi^\nu \partial_\nu E_\mu{}^m + (\partial_\mu \xi^\nu) E_\nu{}^m = \delta_{\rm G}(\xi^\rho) E_\mu{}^m ~~.\end{aligned} \tag{2.6}$$

This is a significant result, because it means that the field $E_\mu{}^m$ transforms under supersymmetry, as if it were a vielbein transforming under the general coordinate transformation $\delta_{\rm G}$ with the parameter $\xi^\mu$ not only in $D=4$ [1] but also in any space-time dimensions! Now the advantage of using the two distinct sets of indices $m, n, \cdots$ and $\mu, \nu, \cdots$ on $E_\mu{}^m$ is also clear: They respectively behave like the curved frame and local Lorentz frame indices.

Once this is understood, it is straightforward to confirm the invariance of the action

$$I_E \equiv \int d^D x \det(E_\mu{}^m) \equiv \int d^D x\, E \equiv \int d^D x\, \mathcal{L}_E ~~, \quad E \equiv \det(E_\mu{}^m) ~~, \tag{2.7}$$

under the variation in (2.1), as a total divergence:[7]

$$\delta_Q \mathcal{L}_E = E E_m{}^\mu \delta_Q E_\mu{}^m = E E_m{}^\mu \left[\xi^\nu \partial_\nu E_\mu{}^m + (\partial_\mu \xi^\nu) E_\nu{}^m\right] = \partial_\mu(E\xi^\mu) ~~. \tag{2.8}$$

We can reconfirm this 'formal' proof by the direct variation with respect to the NG-fermion whose details are to be skipped here.

As the $\lambda$-field equation or the original lagrangian (2.7) reveals, the action $I_E$ contains the standard kinetic term for the NG-fermion, upon the expansion $\det(E_\mu{}^m) \approx 1 + i(\overline{\lambda}\gamma^\mu \partial_\mu \lambda) + \mathcal{O}(\lambda^4)$. Therefore, the NG-fermion is regarded as a 'physical' field, also with self-interaction terms that make our system highly non-trivial.

### 2.2 More Generalizations

As the reader may have already noticed, once the transformation rule of the vielbein is as appears in (2.5), we can write other lagrangian densities constructed from the vielbein. For example, we can write lagrangians in terms of the scalar curvature made up of the 'metric'[8] $G_{\mu\nu} \equiv \eta_{mn} E_\mu{}^m E_\nu{}^n$ with its inverse metric $G^{\mu\nu}$ : $G_{\mu\nu} G^{\nu\rho} = \delta_\mu{}^\rho$, such as the 'Hilbert action':

---

[6]The index for the $N$-extended supersymmetry can be implicit in the underlined indices. We will return to this point later.

[7]The existence of the inverse vielbein $E_m{}^\mu$ is taken for granted otherwise the action is ill-defined.

$$I_R \equiv \int d^D x \sqrt{-G} R(G) \ ,$$

$$R(G) \equiv G^{\mu\nu} R_{\mu\nu}(G) \ , \quad R_{\mu\nu}(G) \equiv R_{\mu\rho\nu}{}^\rho(G) \ ,$$

$$R_{\mu\nu\rho}{}^\sigma(G) \equiv \partial_\mu \left\{{}^{\sigma}_{\nu\rho}\right\} - \partial_\nu \left\{{}^{\sigma}_{\mu\rho}\right\} - \left\{{}^{\tau}_{\mu\rho}\right\}\left\{{}^{\sigma}_{\tau\nu}\right\} + \left\{{}^{\tau}_{\nu\rho}\right\}\left\{{}^{\sigma}_{\tau\mu}\right\} \ ,$$

$$\left\{{}^{\rho}_{\mu\nu}\right\} \equiv \tfrac{1}{2} G^{\rho\sigma}(\partial_\mu G_{\nu\sigma} + \partial_\nu G_{\mu\sigma} - \partial_\sigma G_{\mu\nu}) \ , \tag{2.9}$$

because $G_{\mu\nu}$ transforms as

$$\delta_Q G_{\mu\nu} = \xi^\rho \partial_\rho G_{\mu\nu} + (\partial_\mu \xi^\rho) G_{\rho\nu} + (\partial_\nu \xi^\rho) G_{\mu\rho} = \delta_{\rm G}(\xi^\rho) G_{\mu\nu} \ , \tag{2.10}$$

as if it were a 'metric' tensor. Accordingly, the curvature 'tensors' above transform as

$$\delta_Q R_{\mu\nu\rho}{}^\sigma(G) = \delta_{\rm G}(\xi^\tau) R_{\mu\nu\rho}{}^\sigma(G) \ ,$$

$$\delta_Q R_{\mu\nu}(G) = \delta_{\rm G}(\xi^\tau) R_{\mu\nu}(G) \ , \quad \delta_Q R(G) = \delta_{\rm G}(\xi^\tau) R(G) \ . \tag{2.11}$$

In the case we need more 'canonical' kinetic term for the NG-fermion $\lambda$, we can simply add the action $I_E$ (the analog of a 'cosmological term') to $I_R$, so that a kinetic term for the $\lambda$-field is present, while higher-order terms are regarded as interaction terms. In fact, if the total action is $I_E + I_R$, the lowest-order term in $I_R$ is already at the fourth order in $\lambda$ as an interaction term.

Some readers may wonder if the Riemann or Ricci tensor built out of the metric $G_{\mu\nu} \equiv \eta_{mn} E_\mu{}^m E_\nu{}^n$ vanishes by the use of the $\lambda$-field equation. We show this is not the case, e.g., for the Ricci tensor

$$R_{\mu\nu}(G) = -2i[\overline{\lambda},{}_\rho \gamma^\rho \lambda,{}_{\mu\nu} - \overline{\lambda},{}_{(\mu|}\gamma^\rho \lambda,{}_{|\nu)\rho} - \overline{\lambda},{}_\rho \gamma_{(\mu|}\lambda,{}_{|\nu)}{}^\rho + \overline{\lambda},{}_{(\mu}\gamma_{\nu)}\lambda,{}_\rho{}^\rho] + \mathcal{O}(\lambda^4) \ . \tag{2.12}$$

Here $\lambda,{}_\mu \equiv \partial_\mu \lambda$, $\lambda,{}_{\mu\nu} \equiv \partial_\mu \partial_\nu \lambda$, etc.[9] Obviously, even though $i\gamma^\mu \partial_\mu \lambda \approx \mathcal{O}(\lambda^3)$ by the NG-fermion equation at the lowest order, there still remains non-vanishing component even for the scalar curvature $R(G)$. We can further elaborate our result, introducing more higher-derivative lagrangians, such as $(R_{\mu\nu\rho}{}^\sigma(G))^2$, $(R_{\mu\nu}(G))^2$, or $R(G)^2$, etc.

We stress here that the invariance of $I_E$ or $I_R$ relied only on the particular transformation rule (2.1) of the vielbein, but did not use any property related to the space-time dimensions, such as Fierz arrangements. This implies that the non-linear realizations of supersymmetry should be valid in $\forall D$, where we can define a spinor field with the usual kinetic term. In this context, the specific nature of the spinors (Weyl, Majorana, or Dirac) does not matte.

---

[9]Since only lowest-order terms are under question here, the raising/lowering indices by the metric $G_{\mu\nu}$ or $\eta_{\mu\nu}$ does not matter here.

The generalization to extended supersymmetry is also straightforward, assigning an explicit $N$-supersymmetry index $i$ on $\lambda^i$. Accordingly, (2.1) and (2.2) are now

$$\delta_Q \lambda^i = \epsilon^i + \xi^\mu \partial_\mu \lambda^i \ , \qquad \xi^\mu \equiv i(\bar{\epsilon}^i \gamma^\mu \lambda^i) \ . \tag{2.13}$$

If we suppress the explicit contraction index $i$ as understood, then formally the same equations as (2.4) through (2.9) follow, and there is no restriction on the value of $N$. We have thus a non-linear realization of arbitrarily extended global supersymmetries.

Note that we are dealing with two different space-times, the a 'flat' space-time with no gravity and with no curvature, while the other space-time has non-vanishing curvature (2.12), in which we can define general coordinate transformations. Interestingly, the action $I_E$ is invariant in both of these space-time simultaneously.

Our result applies even to $D \geq 12$ for the invariant actions under non-linear supersymmetry, including of course includes $D = 12$ of F-theory [25] or $D = 13$ of S-theory [26], and so forth, as well as the M(atrix)-theory in $D = 11$ [27]. As a matter of fact, we will see a more explicit link with the M(atrix)-theory [27] in section 3.3.

## 3. Matter Lagrangians

### 3.1 Bosonic Lagrangians

We have so far offered a possible meaning of our 'vielbein', only with the NG-fermion. The next natural question is whether or not we can introduce other matter fields. In the case of $D = 4$, we already know that such lagrangians are easily constructed [18][3][5]. However, because of the latter applications to super $p$-branes, *etc.*, we generalize these into $^\forall D$ and $^\forall N$, even though such a generalization looks straightforward. We show that the answer to this question is in the affirmative, *i.e.*, we can introduce other matter fields, as long as our general covariance $\delta_G$ is maintained.

First, we can show that the algebra of supersymmetry closes on the following scalars $\varphi^a$, a non-Abelian gauge field $A_\mu{}^I$, an $n$-th rank covariant tensor field $T_{\mu_1 \cdots \mu_n}$ with no restriction on symmetry or antisymmetry:

$$\delta_Q \varphi^a = i(\bar{\epsilon}\gamma^\mu \lambda)\partial_\mu \varphi^a = \xi^\mu \partial_\mu \varphi^a = \delta_G(\xi^\rho)\varphi^a \ , \tag{3.1a}$$

$$\delta_Q A_\mu{}^I = \xi^\nu \partial_\nu A_\mu{}^I + (\partial_\mu \xi^\nu) A_\nu{}^I = \delta_G(\xi^\rho) A_\mu{}^I \ , \tag{3.1b}$$

$$\delta_Q T_{\mu_1 \cdots \mu_n} = \xi^\rho \partial_\rho T_{\mu_1 \cdots \mu_n} + (\partial_{\mu_1}\xi^\rho)T_{\rho\mu_2\cdots\mu_n} + \cdots + (\partial_{\mu_n}\xi^\rho)T_{\mu_1\cdots\mu_{n-1}\rho} = \delta_G(\xi^\rho)T_{\mu_1\cdots\mu_n} \ . \tag{3.1c}$$

Here and throughout this paper we use exactly the same $\xi^\mu$ as (2.1). We use the indices $a, b, \cdots$ for an appropriate representation of a certain compact non-Abelian gauge group $G$ for the scalars, while $I, J, \cdots$ for the adjoint representation of $G$. Needless to say, if some of the indices on the tensor are contravariant ones, we can change the signs of terms corresponding to those contravariant indices in (3.1c), following the usual textbook result. Interestingly, the closure on the vector or tensors generates *no* gauge transformations as in (2.3), in addition to the standard translation:

$$[\,\delta_Q(\epsilon_1),\delta_Q(\epsilon_2)\,]\phi = \delta_P(\zeta_{12}^m)\phi \ , \quad (3.2)$$

where $\phi$ is an arbitrary field in (3.1). This is another aspect of the non-linear supersymmetry different from the linear supersymmetry, in which such gauge transformations are induced in the commutator algebras [28].

The construction of some invariant 'matter' actions is now an easy task, as long as our general covariance is maintained *via* $G_{\mu\nu}$ or $E_\mu{}^m$. A typical bosonic action is like

$$I_{\rm B} = \int d^D x \sqrt{-G}\,\left[\,+\tfrac{1}{2}G^{\mu\nu}(D_\mu\varphi^a)(D_\nu\varphi^a) - \tfrac{1}{4}G^{\mu\rho}G^{\nu\sigma}F_{\mu\nu}{}^I F_{\rho\sigma}{}^I\,\right] \ , \quad (3.3)$$

where $D_\mu\varphi^a$ is the usual gauge covariant derivative $D_\mu\varphi^a \equiv \partial_\mu\varphi^a + A_\mu{}^I (T^I)^{ab}\varphi^b$ for an appropriate representation of the generators $T^I$ of the gauge group $G$, and $F_{\mu\nu}{}^I \equiv \partial_\mu A_\nu{}^I - \partial_\nu A_\mu{}^I + f^{IJK} A_\mu{}^J A_\nu{}^K$ is the usual field strength. These gauge covariant combinations transform under $\delta_Q$ as

$$\delta_Q(D_\mu\varphi^a) = \xi^\nu \partial_\nu(D_\mu\varphi^a) + (\partial_\mu\xi^\nu)D_\nu\varphi^a = \delta_{\rm G}(\xi^\rho)(D_\mu\varphi^a) \ , \quad (3.4a)$$

$$\delta_Q F_{\mu\nu}{}^I = \xi^\rho \partial_\rho F_{\mu\nu}{}^I + (\partial_\mu\xi^\rho)F_{\rho\nu}{}^I + (\partial_\nu\xi^\rho)F_{\mu\rho}{}^I = \delta_{\rm G}(\xi^\rho)F_{\mu\nu}{}^I \ . \quad (3.4b)$$

It is not a problem at all to generalize the action (3.3) to more sophisticated interactions. If the above result is only for $D=4$, it is nothing new [18][5][3][20], but our point here is that these actions are valid for $\forall N$-extended supersymmetry in $\forall D$.

### 3.2 Non-Abelian Supersymmetric Born-Infeld Lagrangians

With these relations at hand, it is now straightforward even to supersymmetrize a Born-Infeld lagrangian. This can be easily done, because the combination $H_{\mu\nu} \equiv G_{\mu\nu} + F_{\mu\nu}$ for an Abelian field strength $F_{\mu\nu}$ transforms 'covariantly': $\delta_Q H_{\mu\nu} = \delta_Q(G_{\mu\nu}+F_{\mu\nu}) = \delta_{\rm G}(\xi^\rho)H_{\mu\nu}$, therefore the action[10]

$$I_{\rm SBI} \equiv \int d^D x\,[\,-\det(G_{\mu\nu}+F_{\mu\nu})\,]^{1/2} = \int d^D x\,[\,-\det(H_{\mu\nu})\,]^{1/2} \quad (3.5)$$

is invariant under supersymmetry: $\delta_Q I_{\rm SBI} = 0$. This sort of supersymmetric Born-Infeld lagrangian in particular dimensions, such as in $D=10$ is nothing new [8][9], but our lagrangian is valid for $\forall D$ and $\forall N$.

It is not too difficult to generalize this result to the non-Abelian gauge group. Following the non-supersymmetric non-Abelian Born-Infeld action in [16], we can easily postulate a non-Abelian analog of the supersymmetric Born-Infeld action (3.5), as

$$I_{\rm NASBI} = \int d^D x\,{\rm STr}\,\left[-\det(G_{\mu\nu}I + F_{\mu\nu}{}^I T^I)\right]^{1/2} \equiv \int d^D x\,{\rm STr}\,[\,-\det(H_{\mu\nu})\,]^{1/2} \ , \quad (3.6)$$

---
[10]We need the negative signs in the square roots, because $\det(\eta_{mn}) < 0$.

where the generators $T^I$ are in an appropriate matrix representation, and $I$ in $G_{\mu\nu}I$ is the unit matrix for such a representation. The determinant operation here is only for the indices $\mu\nu$, but not for the group indices. The STr is the symmetrized trace defined by $\text{STr}\,(A_1\cdots A_n) \equiv (1/n!)\,\text{Tr}\,(A_1\cdots A_n + \text{`all permutations'})$. Due to the $\delta_Q$-transformation property

$$\delta_Q H_{\mu\nu} = \delta_Q(G_{\mu\nu}I + F_{\mu\nu}{}^I T^I) = \delta_{\text{G}}(\xi^\rho) H_{\mu\nu} \quad, \tag{3.7}$$

of the combination $H_{\mu\nu} \equiv G_{\mu\nu}I + F_{\mu\nu}{}^I T^I$, together with the property of STr, it is clear that the action $I_{\text{NASBI}}$ is invariant under $\delta_Q$. As before, since $I_{\text{NASBI}}$ is valid for $\forall D$ and $\forall N$, our result is a generalization of [8][9], up to discrepancy in higher-order terms. In fact, it is easy to see that the lowest-order terms in our (3.6) agree with those in [8][9].

Needless to say, we can also add the antisymmetric potential $b_{\mu\nu}$ to (3.5):

$$H_{\mu\nu} = G_{\mu\nu} + F_{\mu\nu} + b_{\mu\nu} \quad, \tag{3.8}$$

while the standard kinetic term for its field strength $G_{\mu\nu\rho} \equiv 3\partial_{[\mu} b_{\nu\rho]}$ is

$$I_b = \int d^D x \,[\, \tfrac{1}{12}\sqrt{-G}\, G^{\mu\sigma} G^{\nu\tau} G^{\rho\upsilon} G_{\mu\nu\rho} G_{\sigma\tau\upsilon}\,] \quad, \tag{3.9}$$

as in super D-branes [13]. In other words, there can be a general antisymmetric component $b_{\mu\nu}$ for the 'metric' $G_{\mu\nu}$, maintaining the $\delta_Q$-invariance as well as the $b$-field gauge invariance: $\delta_\alpha b_{\mu\nu} = 2\partial_{[\mu}\alpha_{\nu]}$. Adding further an appropriate minimal couplings of $A_\mu$ to scalars like $I_{\text{B}}$ (3.3), there is no worry that the field strength $F_{\mu\nu}$ is completely gauged away in (3.8) by the field redefinition of $b_{\mu\nu}$.

For the $N$-extended supersymmetry for $N \geq 2$, we can introduce an $SO(N)$ gauge field $A_\mu{}^{ij}$ ($i, j, \cdots = 1, 2, \cdots, N$) minimally coupled to $\lambda^i$, when all the components of $\lambda$ are Majorana spinors or (Majorana-)Weyl spinors in the same chirality. Such a vector field is also subject to the 'covariant' transformation rule (3.1b). Relevantly, the derivative in (2.5) is replaced by the covariant derivative $D_\mu \lambda^i \equiv \partial_\mu \lambda^i + A_\mu{}^{ij} \lambda^j$:

$$E_\mu{}^m = \delta_\mu{}^m + i(\overline{\lambda}^i \gamma^m D_\mu \lambda^i) \quad. \tag{3.10}$$

One important case is when $i, j, \cdots$ indices are adjoint ones $I, J, \cdots = 1, \cdots, \dim G$ for $D_\mu \lambda^I \equiv \partial_\mu \lambda^I + f^{IJK} A_\mu{}^J \lambda^K$:

$$E_\mu{}^m = \delta_\mu{}^m + i(\overline{\lambda}^I \gamma^m D_\mu \lambda^I) \quad. \tag{3.11}$$

Then the lowest-order terms of our action (3.6) in $D = 10$ coincide with those in linearly supersymmetric Born-Infeld action [9]. This seems to imply that our action (3.6) can be of equal importance for the M(atrix)-theory [27] as the D-brane actions [10][11][12][13].[11]

---

[11]See, however, the recent development about subtlety at higher orders [29]. We do not get into these details in this paper.

Interestingly, the number $N$ of extended supersymmetries in our formulation will be $N = \dim G$, e.g., $N = \mathcal{N}^2$ for $G = U(\mathcal{N})$ used for M(atrix)-theory [27]. Therefore the usual large $\mathcal{N}$ limit for $U(\mathcal{N})$ [27] corresponds to $N = \mathcal{N}^2 \to \infty$, namely infinitely many ($\aleph_0$) extended supersymmetries in our formulation!

### 3.3 Fermionic Lagrangians

We can incorporate other fermionic fields rather easily, because we need only global Lorentz invariance for our action, much like the case of general covariance, which is not really present. In fact, we can show that the action[12]

$$I_{\mathrm{F}} = \int d^D x \tfrac{i}{2} E E_m{}^\mu (\overline{\psi}^a \gamma^m D_\mu \psi^a) \ , \qquad (3.12)$$

for a fermion $\psi^a$ is actually invariant under supersymmetry:

$$\delta_Q \psi^a = i(\overline{\epsilon}\gamma^\mu \lambda)\partial_\mu \psi^a = \xi^\mu \partial_\mu \psi^a = \delta_{\mathrm{G}}(\xi^\mu)\psi^a \ . \qquad (3.13)$$

We use the indices $a, b, \cdots$ for some appropriate representation of the group $G$ that $\psi$ belongs to, and $D_\mu$ is the gauge covariant derivative $D_\mu \psi^a \equiv \partial_\mu \psi^a + A_\mu{}^I (T^I)^{ab}\psi^b$. The invariance $\delta_Q I_{\mathrm{F}} = 0$ is due to the transformations

$$\delta_Q (D_\mu \psi^a) = \xi^\nu \partial_\nu (D_\mu \psi^a) + (\partial_\mu \xi^\nu) D_\nu \psi^a = \delta_{\mathrm{G}}(\xi^\nu)(D_\mu \psi^a) \ , \qquad (3.14)$$

like a 'general covariant vector', with respect to the index $\mu$. As in the case of $\lambda$, there is no essential restriction on the property of the spinor $\psi$, such as Weyl, Majorana, Majorana-Weyl, or Dirac spinor, with any additional group indices in arbitrary $^\forall D$.

Once we have a fermionic field $\psi$, then we can modify (2.5) as

$$E_\mu{}^m = \delta_\mu{}^m + i(\overline{\lambda}\gamma^m \partial_\mu \lambda) + i(\overline{\psi}^a \gamma^m D_\mu \psi^a) \ , \qquad (3.15)$$

This is because $(\overline{\psi}^a \gamma^m D_\mu \psi^a)$ transforms as a 'covariant vector' with respect to the index $\mu$. Accordingly we have the alternative fermionic action

$$I'_{\mathrm{E}} = \int d^D x \det\left[\delta_\mu{}^m + i(\overline{\lambda}\gamma^m \partial_\mu \lambda) + i(\overline{\psi}^a \gamma^m D_\mu \psi^a)\right] \ . \qquad (3.16)$$

We have thus two actions in good contrast, $I'_{\mathrm{E}}$ for fermions, and $I_{\mathrm{NASBI}}$ for bosons.

### 3.4 Chern-Simons Lagrangians

Our results above have a wide range of applications. Since $I_{\mathrm{NASBI}}$ needs no explicit 'metric', but the GN-field forms a composite metric, the first interesting application is the Chern-Simons action for an arbitrary (super)group in $D = 2n+1$ [30][31][32]:

---

[12]In order to avoid any misinterpretation, we always use $E_\mu{}^m$ or $E_m{}^\mu$ explicitly in this paper, *instead of* $\gamma^\mu \equiv \gamma^m E_m{}^\mu$, etc.

$$I_{\text{CS}} = \int_{M^D} \text{STr}\,(\overbrace{F \wedge F \wedge \cdots \wedge F}^{n} \wedge A + \text{`Chern-Simons completion'}) \tag{3.17a}$$

$$= \int_{M^{D+1}} \text{STr}\,(\underbrace{\widehat{F} \wedge \widehat{F} \cdots \wedge \widehat{F}}_{n+1})\ , \tag{3.17b}$$

to be added to the above actions: $I \equiv I_{\text{CS}} + I_{\text{NASBI}} + I'_{\text{E}} + I_{\text{B}}$. Here STr acts on fermionic generators as an antisymmetrized trace, as a supergroup generalization of the symmetrized trace for ordinary groups. In (3.17b), the *hatted* field strength $\widehat{F}_{\hat\mu\hat\nu}$ is defined in the extended space-time $D+1 = 2n+2$ with one additional Vainberg coordinate $0 \leq y \leq 1$ [33][34] with the indices $(\widehat{x}^{\hat\mu}) \equiv (\widehat{x}^0,\ \widehat{x}^1, \cdots,\ \widehat{x}^{2n},\ \widehat{x}^{2n+1} \equiv y) \equiv (x^\mu, y)$ by

$$\widehat{F}_{\hat\mu\hat\nu} \equiv \widehat\partial_{\hat\mu} \widehat{A}_{\hat\nu} - \widehat\partial_{\hat\nu} \widehat{A}_{\hat\mu} + [\widehat{A}_{\hat\mu}, \widehat{A}_{\hat\nu}\} \equiv \begin{cases} \widehat{F}_{\mu\nu} \equiv \partial_\mu \widehat{A}_\nu - \partial_\nu \widehat{A}_\mu + [\widehat{A}_\mu, \widehat{A}_\nu\}\ , \\ \widehat{F}_{y\nu} \equiv \partial_y \widehat{A}_\nu = -\widehat{F}_{\nu y}\ , \end{cases}$$

$$\widehat{A}_\mu(x^\nu, 1) \equiv A_\mu(x^\nu)\ , \qquad \widehat{A}_\mu(x^\nu, 0) = 0\ , \qquad \widehat{A}_y = 0\ , \tag{3.18}$$

in such a way that the leading term (3.17a) is generated by the $y$-integration $\int_0^1 dy$ [33]. The gauge group here can be even a supergroup, such as $OSp(N|m)$ for Chern-Simons supergravity in $^\forall D = 2n+1$ [30][31][32]. Accordingly, the generators $T^I$ in (3.6) should be those of the gauge (super)group under consideration. Due to our transformation rule (3.1b), the invariance $\delta_Q I_{\text{CS}} = 0$ is clear.

The important point is that in the conventional Chern-Simons formulations including those with supergroups [30][31][32], the introduction of additional 'matter' fields has been problematic, because the kinetic terms of such matter fields need some 'metrics' or 'vielbeins'. However, since $I_{\text{NASBI}}$ has no fundamental metric, it is compatible with such formulations without a metric by construction. In this sense, the non-linear realization of supersymmetry is more natural for Chern-Simons theories. Due to the minimal couplings in $I_{\text{F}}$ or $I_{\text{B}}$, there are still non-trivial interactions of the gauge fields to other 'matter' bosonic or fermionic fields.

To put this differently, the important conclusion is that *any* Chern-Simons action (3.17) formulated in $^\forall D = 2n + 1$ for a (super)group is invariant under *hidden* global non-linear supersymmetry dictated by (3.1b). Moreover, once the action $I_E$ (2.7) is added to $I_{\text{CS}}$, then the GN-field becomes physical, and the total action is no longer invariant under arbitrary general coordinate transformations, so that the GN-fermion is no longer gauged away. Therefore the global supersymmetry is not a superficial symmetry any longer by the presence of $I_E$. This has a very far-reaching conclusion, because any Chern-Simons theory for a supergroup [30][31][32] *except in* $D = 3$ [35] has been supposed to have no direct link with the Poincaré supersymmetry with diffeomorphism, in particular in the context of M-theory [31][36].[13] This missing link has been a major drawback with the Chern-Simons formulation [30][31][32] in the context of space-time supersymmetry. In other words, the non-linear representation of supersymmetry may well provide a 'missing link' between Chern-Simons theory for supergroups [30][31][32] and space-time supergravity theories [38] with diffeomorphism.

---

[13]See, however, ref. [37] in which extended objects are introduced to Chern-Simons supergravity to study such a direct link with Poincaré supergravity.

### 3.5 Generalized $\sigma$-Model Lagrangians

We can try even a more sophisticated example. Suppose we have a coset $G/H$, whose coordinates are parametrized by the scalars $\varphi^a$. Now instead of $H_{\mu\nu}$ in (3.5), we consider

$$H_{\mu\nu} \equiv G_{\mu\nu} + F_{\mu\nu} + g_{ab}(\varphi)(\partial_\mu \varphi^a)(\partial_\nu \varphi^b) \ , \tag{3.19}$$

where $g_{ab}(\varphi)$ is the metric on this coset. Due to the transformation property of $\partial_\mu \varphi^a$ under $\delta_Q$, it is clear that the action

$$I_\sigma = \int d^D x \left[ -\det \left\{ G_{\mu\nu} + F_{\mu\nu} + g_{ab}(\varphi)(\partial_\mu \varphi^a)(\partial_\nu \varphi^b) \right\} \right]^{1/2} \ , \tag{3.20}$$

is invariant under $\delta_Q$. As we easily see, the lowest order-term in the determinant contains the same kinetic term for the scalars $\varphi^a$ as the conventional $\sigma$-model. We can further add some Wess-Zumino-Novikov-Witten (WZNW) term to $I_\sigma$, as an analog of WZNW-terms used in D-brane theories [13][9][11].

### 3.6 Duality Transformation of $F_{\mu\nu}$ into $N_{\mu_1 \cdots \mu_{D-2}}$

As another interesting by-product, we mention the duality transformation [39] for the field strength $F_{\mu\nu}$ into $N_{\mu_1 \cdots \mu_{D-2}}$ in $^\forall D$. We can show this is not too difficult, even with the general 'metric tensor' like that in (3.20). Such a duality transformation has been done in $D = 4$ [40], but our formulation here is complimentary, applicable to $^\forall D \neq 4$ and $^\forall N$.

We start with the generalized supersymmetric Born-Infeld action

$$I'_{\text{SBI}} \equiv \int d^D x \left[ -\det(g_{\mu\nu} + F_{\mu\nu}) \right]^{1/2} \ , \tag{3.21}$$

for an Abelian field strength $F_{\mu\nu}$. Here $g_{\mu\nu}$ can be any generalized metric containing any fields other than $A_\mu$ or $F_{\mu\nu}$, e.g.,

$$g_{\mu\nu} \equiv G_{\mu\nu} + g_{ab}(\varphi)(\partial_\mu \varphi^a)(\partial_\nu \varphi^b) \ , \tag{3.22}$$

in the case of (3.20). The crucial point here is that $g_{\mu\nu}$ should have *no* coupling to $A_\mu$ for the duality transformation [39] to work.

The lagrangian (3.21) is not very convenient for duality transformations, because the lowest-order term in $F$ is not manifestly quadratic in $F$, is not manifest. In order to make the $F^2$-term more manifest, we first introduce the generalized vielbein

$$g_{\mu\nu} = e_\mu{}^m \eta_{mn} e_\nu{}^n \ . \tag{3.23}$$

Note that this $e_\mu{}^m$ is more generalized than (2.5), and it contains at least the latter as a part of it. Using such a vielbein, we see that

$$\det(g_{\mu\nu} + F_{\mu\nu}) = \det[e_\mu{}^m(\delta_m{}^n + F_m{}^n)e_{\nu n}]$$
$$= (\det e_\mu{}^m)(\det e_{\nu n})\det(\delta_m{}^n + F_m{}^n) = \det(g_{\mu\nu})\det(\delta_m{}^n + F_m{}^n) , \quad (3.24)$$

where as usual $F_{mn} \equiv e_m{}^\mu e_n{}^\nu F_{\mu\nu}$, $F_m{}^n \equiv \eta^{nr} F_{mr}$, etc. Note also that the antisymmetry $F^t = -F$ implies that

$$\det(\delta_m{}^n + F_m{}^n) = \det(I + F) = \det\{(I+F)^t\} = \det(I - F)$$
$$= [\det(I+F)\det(I-F)]^{1/2} = [\det\{(I+F)(I-F)\}]^{1/2}$$
$$= [\det(I - F^2)]^{1/2} = [\det\{\delta_m{}^n - (F^2)_m{}^n\}]^{1/2} , \quad (3.25)$$

with $(F^2)_m{}^n \equiv F_m{}^r F_r{}^n$. Hence (3.21) is equivalent to

$$I'_{\text{SBI}} = \int d^D x \sqrt{-g}\, [\det\{\delta_m{}^n - (F^2)_m{}^n\}]^{1/4} , \quad (3.26)$$

where $g \equiv \det(g_{\mu\nu})$ and the $F^2$-term is now manifest.

We next 'linearize' the determinant operation, in such a way that the $F^2$-term is involved only as a quadratic term, by introducing a symmetric matrix auxiliary field $P_m{}^n$:

$$I'_{\text{SBI}} = \int d^D x\, a\sqrt{-g}\left[cP^{1/4} + P^{1/4}(P^{-1})_m{}^n\{\delta_n{}^m - (F^2)_n{}^m\}\right] , \quad (3.27)$$

where $P \equiv \det(P_m{}^n)$, and the constants $a$ and $c$ are

$$a = \tfrac{1}{4} , \quad c = 4 - D , \quad (3.28)$$

in order to get the normalized algebraic field equation for $P_m{}^n$:

$$P_m{}^n = \delta_m{}^n - (F^2)_m{}^n . \quad (3.29)$$

Here the matrix multiplication for $F^2$ is done with the local Lorentz indices $m, n, \cdots$.

By these preliminaries, it is now easy to perform the duality transformation [39] from $F_{\mu\nu}$ into $N_{\mu_1\cdots\mu_{D-2}}$. As usual, we introduce the lagrange multiplier potential $M_{\mu_1\cdots\mu_{D-3}}$ into a constraint action, so that the total action is now

$$I''_{\text{SBI}} = \int d^D x \left[ ac\sqrt{-g}P^{1/4} + a\sqrt{-g}P^{1/4}(P^{-1})_m{}^n\{\delta_n{}^m - (\widehat{F}^2)_n{}^m\} \right.$$
$$\left. + \tfrac{2}{(D-1)!}\epsilon^{\nu\rho\sigma\mu_1\cdots\mu_{D-3}} M_{\mu_1\cdots\mu_{D-3}} \partial_\nu \widehat{F}_{\rho\sigma} \right] \quad (3.30a)$$
$$= \int d^D x \left[ ac\sqrt{-g}P^{1/4} + a\sqrt{-g}P^{1/4}(P^{-1})_m{}^n\{\delta_n{}^m - (\widehat{F}^2)_n{}^m\} \right.$$
$$\left. - \tfrac{2}{(D-2)!}\epsilon^{\mu_1\cdots\mu_{D-2}\rho\sigma} N_{\mu_1\cdots\mu_{D-2}} \widehat{F}_{\rho\sigma} \right] . \quad (3.30b)$$

Here $\widehat{F}_{\mu\nu}$ is no longer a field strength of $A_\mu$, but is regarded as an independent field variable [39], while $N_{\mu_1\cdots\mu_{D-2}} \equiv (D-2)\partial_{[\mu_1} M_{\mu_2\cdots\mu_{D-2}]}$. The constraint term at the end of (3.30a) forces the Bianchi identity for $\widehat{F}_{\mu\nu}$ for consistency. Our next step is to obtain the algebraic field equation of $\widehat{F}_{\mu\nu}$, as

$$\frac{\delta}{\delta \widehat{F}_{\mu\nu}} I''_{\text{SBI}} = Q^{\mu\rho}\widehat{F}_\rho{}^\nu + \widehat{F}^{\mu\rho}Q_\rho{}^\nu - 2\widetilde{N}^{\mu\nu} = 0 ~, \tag{3.31}$$

where

$$Q_\mu{}^\nu \equiv P^{1/4}(P^{-1})_\mu{}^\nu \equiv P^{1/4} e_\mu{}^m e_n{}^\nu (P^{-1})_m{}^n ~, \tag{3.32}$$

and

$$\widetilde{N}^{\mu\nu} \equiv \frac{1}{\sqrt{-g}} \frac{1}{(D-2)!} \epsilon^{\mu\nu\rho_1\cdots\rho_{D-2}} N_{\rho_1\cdots\rho_{D-2}} ~. \tag{3.33}$$

The field equation (3.31) is conveniently expressed as a $D \times D$ matrix equation

$$\{Q, \widehat{F}\}_m{}^n = 2\widetilde{N}_m{}^n ~. \tag{3.34}$$

In order to solve (3.34) for $\widehat{F}$ in terms of $P$ and $\widetilde{N}$, we split $Q$ into $Q = I - 2q$, and get the solution as an infinite series in terms of $q$:

$$\widehat{F} = \sum_{r=0}^\infty \sum_{s=0}^r \binom{r}{s} q^s \widetilde{N} q^{r-s} = \sum_{r=0}^\infty \sum_{s=0}^r \frac{1}{2^r} \binom{r}{s} (I-Q)^s \widetilde{N} (I-Q)^{r-s} ~. \tag{3.35}$$

Even though this is an infinite series, it is a closed form, starting with the lowest-order term $\widehat{F}^{\mu\nu} = \widetilde{N}^{\mu\nu} + \mathcal{O}(q)$, as desired. Our final step is to substitute (3.35) into (3.30b), in order to eliminate $\widehat{F}$. After this, all the $\widehat{F}$-terms are replaced by its dual $N_{\mu_1\cdots\mu_{D-2}}$, and thus we have achieved the duality transformation on a generalized metric (3.23).

Eq. (3.28) indicates that the case of $D=4$ is exceptional or singular. This is because the original field strength and its Hodge-dual have the same rank. This seems also related to the possible $SL(2,\mathbb{R})$ duality invariance between these two field strengths [40]. Other than this exceptional case of $D=4$, our action is still superinvariant, because of the 'general coordinate invariance' valid in $^\forall D \neq 4$ and $^\forall N$.

## 4. Superspace Formulations

### 4.1 Second and Third-Rank Superfield Strengths

Once we have understood the component formulation of the non-linear realization of supersymmetry, our next question is how to re-formulate the same results in superspace [41]. This is a non-trivial question, because even though we know that such a formulation is definitely possible in $D=4$ with auxiliary fields [17], or without auxiliary fields in $D=10$ [9], $D=6$ or $D=7$ [12], it is no longer trivial that we can repeat it in *arbitrary* space-time dimensions $^\forall D$. In particular, it is no longer automatic that superspace Bianchi identities [41] are satisfied, when supersymmetry is realized non-linearly. This is due to possible 'extra' symmetries present in the system, that prevent us from formulating the component results in superspace [35].

We answer this question, by the actual investigation of conventional superspace Bianchi identities [41]. As the first non-trivial example, we study the Bianchi identity[14]

$$\nabla_{[A} F_{BC)}{}^I - T_{[AB|}{}^D F_{D|C)}{}^I \equiv 0 \ , \tag{4.1}$$

for the superfield strength $F_{AB}{}^I$ of a non-Abelian gauge vector superpotential $A_A{}^I$, corresponding to (3.1b).

Our superspace constraints are summarized as

$$T_{\underline{\alpha}\underline{\beta}}{}^c = -2i(\gamma^c)_{\underline{\alpha}\underline{\beta}} \ , \tag{4.2a}$$

$$\nabla_{\underline{\alpha}} \lambda^{\underline{\beta}} = -\delta_{\underline{\alpha}}{}^{\underline{\beta}} + i(\gamma^c)_{\underline{\alpha}\underline{\gamma}} \lambda^{\underline{\gamma}} \nabla_c \lambda^{\underline{\beta}} \equiv -\delta_{\underline{\alpha}}{}^{\underline{\beta}} - i(\gamma^c \lambda)_{\underline{\alpha}} \nabla_c \lambda^{\underline{\beta}} \ , \tag{4.2b}$$

$$F_{\underline{\alpha} b}{}^I = +i(\gamma^c)_{\underline{\beta}\underline{\gamma}} \lambda^{\underline{\gamma}} F_{cb}{}^I \equiv -i(\gamma^c \lambda)_{\underline{\alpha}} F_{cb}{}^I = -P_{\underline{\alpha}}{}^a F_{ab}{}^I \ , \tag{4.2c}$$

$$F_{\underline{\alpha}\underline{\beta}}{}^I = -\tfrac{1}{2} (\gamma^a \lambda)_{(\underline{\alpha}} (\gamma^a \lambda)_{\underline{\beta})} F_{ab}{}^I = +P_{\underline{\alpha}}{}^a P_{\underline{\beta}}{}^b F_{ab}{}^I \ , \qquad P_{\underline{\alpha}}{}^b \equiv i(\gamma^b \lambda)_{\underline{\alpha}} \ . \tag{4.2d}$$

There is no *a priori* restriction on $D$ here.[15] The most important constraint is (4.2d), which is similar to the postulates given in [42][8] in $D=10$ as a special case of $D$. This is because $F_{\underline{\alpha}\underline{\beta}}{}^I$ can be rewritten as

$$F_{\alpha\beta}{}^I = -\tfrac{1}{16}(\gamma_c)_{\alpha\beta} (\overline\lambda \gamma^{abc} \lambda) F_{ab}{}^I + \tfrac{1}{1920} (\gamma_{c_1 \cdots c_5})_{\alpha\beta} (\overline\lambda \gamma^a \gamma^{c_1 \cdots c_5} \gamma^b \lambda) F_{ab}{}^I \ , \tag{4.3}$$

in terms of undotted chiral indices in $D=10$ after Fierzing, where the second term has been suggested in [42][8]. For an arbitrary space-time dimension $D$, we keep the original form (4.2d), because forms like (4.3) needs Fierzing depending on $D$.

The confirmation of Bianchi identities (4.1) goes in a way similar to the conventional case, once we know the structure of $F_{\underline{\alpha}\underline{\beta}}{}^I$. However, there are also important differences. For example, it turns out to be easier to proceed the conformation of Bianchi identities from the highest mass dimension two ($d=2$) to the lowest $d=1/2$ backward, as opposed to the conventional case, where the lowest-dimensional one is the simplest. The reason for this is that the lowest dimensional constraint (4.2c) is more involved than the higher ones. Another important aspect of this system is that the Bianchi identities with these constraints do not yield any field equation for the NG-fermion $\lambda$. This is natural, because as we saw in component results, the transformation rule for $\lambda$ is independent of the choice of the total action, e.g., $I_E$ versus $I_E + I_R$, etc., namely the system of non-linear supersymmetry is essentially 'off-shell'.

---

[14]In this section of superspace, we use the indices $A=(a,\underline\alpha)$, $B=(b,\underline\beta)$, $\cdots$ for superspace coordinates. The underlined spinorial indices $\underline\alpha, \underline\beta, \cdots$ include any possible implicit indices such as 'dottedness' for chiralities, or some $USp(N)$ indices needed for extended supersymmetry, depending on $D$ [24]. Our symmetrizations in this section are not normalized, i.e., $M_{[A} N_{B)} \equiv M_A \mp N_B$.

[15]Depending on $D$, we have to switch between the antisymmetry and symmetry of the metric tensor for spinorial indices [24]. For such cases, we need to switch some signs for contracted indices in the constraints here, but the basic structure is still universal.

Another instructive example is the third-rank superfield strength $G_{ABC}$, satisfying the Bianchi identity

$$\tfrac{1}{6}\nabla_{[A}G_{BCD)} - \tfrac{1}{4}T_{[AB|}{}^{E}G_{E|CD)} \equiv 0 \ . \tag{4.4}$$

Our superspace constraints are

$$G_{\underline{\alpha}bc} = -P_{\underline{\alpha}}{}^{a}G_{abc} \ , \quad G_{\underline{\alpha}\underline{\beta}c} = -i(\gamma_c)_{\underline{\alpha}\underline{\beta}} + P_{\underline{\alpha}}{}^{a}P_{\underline{\beta}}{}^{b}G_{abc} \ , \tag{4.5a}$$

$$G_{\underline{\alpha}\underline{\beta}\underline{\gamma}} = -P_{\underline{\alpha}}{}^{a}P_{\underline{\beta}}{}^{b}P_{\underline{\gamma}}{}^{c}G_{abc} \ , \tag{4.5b}$$

in addition to (4.2a) and (4.2b).

Note the interesting patterns of involvement of the factor $P_{\underline{\alpha}}{}^{b}$ in (4.5) as well as (4.2), namely all the spinorial index $\underline{\alpha}, \underline{\beta}, \cdots$ in the superfields are converted into the bosonic ones $a, b, \cdots$ by the factor $P_{\underline{\alpha}}{}^{a}$. By reading these patterns, it is easy to generalize these results to more higher-rank superfield strengths. The role of such a factor has been noted, e.g., in $D=6$ [12], but our point is that our constraints are valid for $\forall D$ and $\forall N$. Even though we skip the details for these higher-rank superfield strengths in this paper, the geometrical structure of this superspace formulation is much clearer than component formulation. In the next subsection, we address ourselves to the question of coupling to curved superspace backgrounds with supergravity, based on this particular role for the factor $P_{\underline{\alpha}}{}^{b}$.

### 4.2 Super $(2k-1)$-Brane Interpretations[16]

As the pattern of the factor $P_{\underline{\alpha}}{}^{a}$ suggests, all of the effects of the GN-fermions can be re-interpreted as part of the vielbein superfields. For example, the first form (2.5) of $E_\mu{}^m$ suggests that the GN-fermion can be interpreted as nothing else than the $\Theta$-coordinates in *flat* superspace, as has been also pointed out by several authors [13][12][9][18] for special cases of $D=10$ [9], $D=6$, $D=7$ [12], or $D=4$ [18]. In our paper, we generalize this feature to more general dimensions, because this process seems imperative, if we also want to introduce 'curved' backgrounds with supergravity, and we need to consider more geometrical superspace formulation.

If this scenario really works in more general dimensions, we must have the geometric structure of a base manifold and a target superspace at the same time, as special cases [9][12] indicate. Accordingly, we must have an analog of Green-Schwarz type super $p$-brane action [21] in which the base manifold (world-supervolume) and the bosonic coordinates in the target superspace have the same dimensions, with appropriate so-called fermionic $\kappa$-symmetry [43][21]. Such a formulation can accommodate 'curved' superspace supergravity backgrounds, whose classical superfield equations can be satisfied, only when the action has the $\kappa$-invariance. The subtlety here is that there are two bosonic coordinates involved both for the world-supervolume and the target superspace with the same dimensionality, so that we need to avoid the confusion between them. This spirit of having the same dimensionality both for the world-supervolume and the target space-time was also presented in the context of doubly-supersymmetric $p$-branes or L-branes [44], or superembeddings [45][14].

There has already been a general formulation for such a Green-Schwarz $\sigma$-model action in general even dimensions $D = 2k$ in [22] in which the $D = 2k$-th rank antisymmetric potential superfield $C_{A_1 \cdots A_D}$ is introduced, as one more generalization of super $p$-branes in [21], and its $\kappa$-invariance has been confirmed. As a matter of fact, the generalized Volkov-Akulov action $I_E$ (2.7) with the determinant form of $E_\mu{}^m$ already suggests the involvement of such maximally-ranked $C_{A_1 \cdots A_D}$-superfield. In ref. [22], we have seen that a similar putative Green-Schwarz action in odd dimensions $D = 2k - 1$ instead of $D = 2k$ leads to a trivial $\kappa$-symmetry, due to the absence of chirality projection $\gamma$-matrix like $\gamma_5$.

Following the results in [22], our $(2k-1)$-brane action in general $D = 2k$-dimensions should be

$$S_{\mathrm{GS}} = \int d^{2k}\sigma \Big[ + \tfrac{1}{2} \sqrt{-g}\, g^{ij} \eta_{ab} \Pi_i{}^a \Pi_j{}^b - (k-1)\sqrt{-g} + \tfrac{1}{(2k)!} \epsilon^{i_1 \cdots i_{2k}} \Pi_{i_1}{}^{A_1} \cdots \Pi_{i_{2k}}{}^{A_{2k}} C_{A_{2k} \cdots A_1} \Big] \ , \tag{4.6}$$

where $i, j, \cdots = 0, 1, 2, \cdots, D-1 = 2k-1$ are the curved indices for the world-supervolume with the coordinates $\sigma^i$ and the metric $g_{ij}$, while $\Pi_i{}^A \equiv (\partial_i Z^M) E_M{}^A$ are the ordinary pullbacks from the target superspace with the coordinates $(Z^M) \equiv (X^m, \Theta^{\underline{\mu}})$ into the world-supervolume. The superpotential $C_{A_1 \cdots A_{2k}}$ has its superfield strength $H_{A_1 \cdots A_{2k+1}}$, satisfying the Bianchi identities

$$\tfrac{1}{(2k+1)!} \nabla_{[A_1} H_{A_2 \cdots A_{2k+2})} - \tfrac{1}{2[(2k)!]} T_{[A_1 A_2|}{}^B H_{B|A_3 \cdots A_{2k+2})} \equiv 0 \ , \tag{4.7}$$

while the supertorsion satisfies the $T$-Bianchi identities [41]

$$\nabla_{[A} T_{BC)}{}^D - T_{[AB|}{}^E T_{E|C)}{}^D - \tfrac{1}{2} R_{[AB|e}{}^f (\mathcal{M}_f{}^e)_{|C)}{}^D \equiv 0 \ . \tag{4.8}$$

The $H$-Bianchi identities (4.7) at mass dimensions $d = 1/2$ and $1$ [22] require the conditions:

$$T_{(\underline{\alpha\beta}|}{}^{\underline{\epsilon}} H_{\underline{\epsilon}|\underline{\gamma})d_1 \cdots d_{D-1}} = 0 \ , \quad T_{[c_1|(\underline{\alpha}|}{}^{\underline{\delta}} H_{\underline{\delta}|\underline{\beta})|c_2 \cdots c_D]} = 0 \ . \tag{4.9}$$

There are of course other Bianchi identities satisfied by other superfield strength, depending on the supergravity multiplet for a given $D$ and $N$, e.g., the fourth-rank superfield strength $F_{ABC}$ for $D = 10$, $N = 2A$, that we do not present here explicitly.

The superspace constraints at mass dimension $d = 0$ relevant to our $\kappa$-transformation below are

$$T_{\underline{\alpha\beta}}{}^c = -2i(\gamma^c)_{\underline{\alpha\beta}} \quad , \tag{4.10a}$$

$$H_{\underline{\alpha\beta}c_1\cdots c_{2k-1}} = +i(\gamma_{c_1\cdots c_{2k-1}}\gamma_{2k+1})_{\underline{\alpha\beta}} \quad , \tag{4.10b}$$

where $\gamma_{2k+1}$ is an analog of $\gamma_5$ in $D = 4$, and all other independent components in $H_{A_1\cdots A_{2k+1}}$ are zero [22].

Even though our superspace constraints (4.10) look so simple, we can accommodate all the curved supergravity effects in the system, such as the gravitino $E_a{}^{\underline{\beta}} \approx \psi_a{}^{\underline{\beta}}$, as long as the $H$- and $T$-Bianchi identities above, as well as those for other superfield strengths are satisfied, with curved supertorsions, supercurvatures and superfield strengths.

As an important note to avoid confusion, we no longer use the flat case constraints (4.2b) - (4.2d), (4.5), once we have introduced the Green-Schwarz action. They will be recovered, though, when we consider the special flat superspace limit we mention shortly.

It is easy to see that $S_\text{GS}$ has the $\kappa$-invariance under

$$\delta_\kappa E^{\underline{\alpha}} = \tfrac{1}{2}(I+\Gamma)^{\underline{\alpha}}{}_{\underline{\beta}}\kappa^{\underline{\beta}} \quad , \quad \delta_\kappa E^a = 0 \quad , \tag{4.11a}$$

$$\Gamma^{\underline{\alpha}}{}_{\underline{\beta}} \equiv \frac{(-1)^{(k-1)(2k+1)/2}}{(2k)!\sqrt{-g}}\epsilon^{i_1\cdots i_{2k}}\Pi_{i_1}{}^{a_1}\cdots\Pi_{i_{2k}}{}^{a_{2k}}(\gamma_{a_1\cdots a_{2k}})^{\underline{\alpha}}{}_{\underline{\beta}} = (\gamma_{2k+1})^{\underline{\alpha}}{}_{\underline{\beta}} \quad , \tag{4.11b}$$

following ref. [22], whose details we do not repeat here. As in [22], the field equation of $g_{ij}$

$$g_{ij} = \eta_{ab}\Pi_i{}^a\Pi_j{}^b \quad , \tag{4.12}$$

is algebraic, so we do not have to worry about its variation under $\delta_\kappa$.

In order to go back to 'flat' superspace, we perform the identifications $\Theta^{\underline{\alpha}} \equiv \lambda^{\underline{\alpha}}$, $X^m \equiv \sigma^i$,[17] so that $S_\text{GS}$ coincides with the generalized Volkov-Akulov action (2.7), after eliminating $g_{ij}$ by its algebraic field equation (4.12). In fact, the $\Pi$'s coincides with the $E$'s in (2.5):

$$\Pi_i{}^a \rightarrow \delta_i{}^a + i(\overline{\Theta}\gamma^a\partial_i\Theta) \equiv \delta_i{}^a + i(\overline{\lambda}\gamma^a\partial_i\lambda) \quad , \quad \Pi_i{}^{\underline{\alpha}} \rightarrow \partial_i\Theta^{\underline{\alpha}} \equiv \partial_i\lambda^{\underline{\alpha}} \quad . \tag{4.13}$$

In other words, in the flat target superspace, the fermionic coordinates are nothing else than the GN-fermions [9][12]. On the other hand, the $\kappa$-transformation stays formally the same as (4.11), except that we no longer have e.g., the gravitino in the target superspace: $E_a{}^{\underline{\beta}}| \approx \psi_a{}^{\underline{\beta}} \rightarrow 0$. Accordingly, our previous flat space constraints (4.2b) - (4.2d) or (4.5) will be recovered, due to the identification $\Theta \approx \lambda$ in the superspace vielbeins $E_m{}^a$ [41]. Finally, the WZNW-term in (4.6) is shown to be equivalent to $\sqrt{-g}$, as follows: First, we can identify $C_{a_1\cdots a_{2k}} \approx \epsilon_{a_1\cdots a_{2k}}C(Z)$ with a scalar superfield $C(Z)$. Second, due to $H_{\underline{\alpha}b_1\cdots b_{2k}} = 0$, we can set $C(Z) =$ const. Hence the WZNW-term is proportional to $[(2k)!]^{-1}\epsilon^{i_1\cdots i_{2k}}\epsilon_{a_1\cdots a_{2k}}\Pi_{i_1}{}^{a_1}\cdots\Pi_{i_{2k}}{}^{a_{2k}} = \det(\Pi_i{}^a) = \sqrt{-g}$.

---

[17]This identification makes sense, because the dimension of the $\sigma^i$'s and that of $X^m$'s coincide.

The introduction of the $U(1)$ field strength $F_{ij} \equiv \partial_i A_j - \partial_j A_i$ with the fundamental vector $A_i$ on the world-supervolume is not difficult. For this purpose, we modify $S_{\text{GS}}$ as

$$S'_{\text{GS}} = \int d^{2k}\sigma \left[ \tfrac{1}{2}\sqrt{-\widetilde{g}}\widetilde{g}^{ij}(\eta_{ab}\Pi_i{}^a\Pi_j{}^b + F_{ij}) - (k-1)\sqrt{-\widetilde{g}} \right. $$
$$\left. + \tfrac{1}{(2k)!}\epsilon^{i_1\cdots i_{2k}}\Pi_{i_1}{}^{A_1}\cdots\Pi_{i_{2k}}{}^{A_{2k}} C_{A_{2k}\cdots A_1} \right] \ , \tag{4.14}$$

where $\widetilde{g}_{ij}$ is an auxiliary field both with symmetric and antisymmetric components, playing a role of a 'generalized' metric, and $\widetilde{g}^{ij}$ is its inverse with $\widetilde{g} \equiv \det(\widetilde{g}_{ij})$. Since the field equation of $\widetilde{g}_{ij}$ is algebraic:

$$\widetilde{g}_{ij} = \eta_{ab}\Pi_i{}^a\Pi_j{}^b + F_{ij} \ , \tag{4.15}$$

we can eliminate $\widetilde{g}_{ij}$ to get the D-p-brane type action [13][9][12][14]

$$S'_{\text{GS}} \to \int d^{2k}\sigma \left[ \{-\det(\eta_{ab}\Pi_i{}^a\Pi_j{}^b + F_{ij})\}^{1/2} + \tfrac{1}{(2k)!}\epsilon^{i_1\cdots i_{2k}}\Pi_{i_1}{}^{A_1}\cdots\Pi_{i_{2k}}{}^{A_{2k}} C_{A_{2k}\cdots A_1} \right] \ , \tag{4.16}$$

also with the last WZNW-term. The $S'_{\text{GS}}$ has the $\kappa$-invariance under (4.11) and $\delta_\kappa A_i = 0$, as is easily confirmed. We repeat here that these actions are valid not only in $D=10$ like type IIA or IIB superstrings [11][12][9][14], but also in any arbitrary even dimensions $D = 2k$ where we can build supergravity backgrounds consistently in superspace [22].[18]

In this section, we have shown how to accommodate curved supergravity backgrounds into our system, and in particular, we have seen how the generalized Volkov-Akulov action (2.7) can be interpreted as a special case of Green-Schwarz super $(2k-1)$-brane action in general $D = 2k$ dimensions. At the point of introducing supergravity backgrounds, we lose the universality on $D$, due to the satisfaction of $T$-Bianchi identities, which needs consistent supergravity constraints. To put this differently, we have seen in this section that, when the background superspace is flat, supersymmetry is global and non-linear, with no restriction on $D$ or $N$, and we can always define Green-Schwarz super $p$-brane actions. However, when superspace is curved with supergravity with Lorentz covariance, there is a restriction, such as $D \leq 11$ by the ordinary Bianchi identities for the curved backgrounds.

### 4.3 Simplified (Supersymmetry)$^2$-Models

The degeneracy of the dimensionalities between the target space-time and the world-supervolume suggests some formulation similar to what is called (Supersymmetry)$^2$-models [23]. Namely we expect two kinds of supersymmetries realized simultaneously in the target space-time and the world-supervolume.[19] In this sub-section, we show that such a formulation indeed makes sense in general superspace, with the simplification that these two supersymmetries are realized within the same space-time dimensions. Such a sharing of the common space-time is possible, because of the degeneracy of the dimensionalities between the base and target space-time. Accordingly, we have only one set of superspace coordinates $(Z^M) \equiv (x^m, \theta^{\underline{\mu}})$, which we temporarily call 'simplified' (Supersymmetry)$^2$-model.[20]

---

[18]This allows us to even beyond $D = 2k \geq 12$ [46], if we can give up Lorentz covariance.

[19]Similar formulation was also presented for L-brane models [44], in which various off-shell field strengths dual to scalars or vectors on the world-supervolume are introduced.

Our starting point is to postulate the superspace analog of our component vielbein (2.5) by

$$E_M{}^a = E_M^{(0)a} + (-1)^M i(\overline{\Lambda}\gamma^a \partial_M \Lambda) \equiv E_M^{(0)a} + (-1)^M i\Lambda^{\underline{\beta}}(\gamma^a)_{\underline{\beta}}{}^{\underline{\gamma}}\partial_M \Lambda_{\underline{\gamma}} \ , \tag{4.17a}$$

$$E_M{}^{\underline{\alpha}} = E_M^{(0)\underline{\alpha}} = \delta_M{}^{\underline{\alpha}} \ . \tag{4.17b}$$

Here $E_M^{(0)A}$ is the ordinary flat superspace vielbein [41]:

$$(E_M^{(0)A}) \equiv \begin{pmatrix} \delta_m{}^a & O \\ -i(\gamma^a\theta)_{\underline{\mu}} & \delta_{\underline{\mu}}{}^{\underline{\alpha}} \end{pmatrix} \ , \quad (E_A^{(0)M}) \equiv \begin{pmatrix} \delta_a{}^m & O \\ +i(\gamma^m\theta)_{\underline{\alpha}} & \delta_{\underline{\alpha}}{}^{\underline{\mu}} \end{pmatrix} \ . \tag{4.18}$$

The form for $E_M{}^a$ in (4.17a) is a simple generalization of (2.5) into superspace, while there might be an additional term like $\partial_M \Lambda^{\underline{\alpha}}$ in (4.17b), which is excluded here just for simplicity. The Grassmann parity $(-1)^M$ is needed in (4.17a), because of the contracted spinorial indices on $\Lambda$. The $\Lambda^{\underline{\alpha}}$ in (4.17) is a spinor superfield, whose non-linear supersymmetry transformation rule is

$$\delta_Q \Lambda^{\underline{\alpha}} = \epsilon^{\underline{\alpha}} + \xi^M \partial_M \Lambda^{\underline{\alpha}} \ , \tag{4.19}$$

where

$$\xi^m = i(\overline{\epsilon}\gamma^m \Lambda) \ , \quad \xi^{\underline{\mu}} = 0 \ . \tag{4.20}$$

Here as in component case, $\gamma^m \equiv \delta_a{}^m \gamma^a$, while we avoid the usage of $\gamma^m \equiv E_a{}^m \gamma^a$.

We can also obtain the inverse matrix components perturbatively for $E_A{}^M$, as

$$E_A{}^m = E_A^{(0)m} - (-1)^A i(\overline{\Lambda}\gamma^m D_A \Lambda) + \mathcal{O}(\Lambda^4) \ , \tag{4.21a}$$

$$E_A{}^{\underline{\mu}} = \delta_A{}^{\underline{\mu}} \ , \tag{4.21b}$$

up to $\mathcal{O}(\Lambda^4)$-terms ignored compared with the $\mathcal{O}(\Lambda^0)$-terms. Note the absence of $\Lambda$-dependent terms in (4.21b), which is confirmed to all orders by the general method to get an inverse of a supermatrix [38].

We can easily show that (4.19) induces the supersymmetry transformation of the vielbeins

$$\delta_Q E_M{}^A = \xi^N \partial_N E_M{}^A + (\partial_M \xi^N) E_N{}^A \ , \quad \delta_Q E_A{}^M = \xi^N \partial_N E_A{}^M - E_A{}^N (\partial_N \xi^M) \ , \tag{4.22}$$

---

[20]We can in principle elaborate this formulation by introducing the second set of coordinates $(x'^m, \theta'^{\underline{\mu}})$, in order to distinguish the base and target space-times, but we skip such a formulation in this paper.

due to the restriction (4.20). Here we can confirm (4.22b) only up to $\mathcal{O}(\Lambda^3)$-terms because of the ignored terms in (4.21). Eq. (4.22) implies nothing else than the general coordinate transformation of the vielbein in superspace [41].

When we consider some other superfields, such as a real scalar superfield $\Phi$, its supersymmetry transformation rule is

$$\delta_Q \Phi = \xi^N \partial_N \Phi = \delta_G(\xi^M)\Phi \quad , \tag{4.23}$$

to comply with (4.21). It is not difficult to show that the commutator of two such non-linear supersymmetries on $\Lambda^{\underline{\alpha}}$ or $\Phi$ induces the desirable translations:

$$[\delta_Q(\epsilon_1), \delta_Q(\epsilon_2)] = +\delta_P(\zeta_{12}^a) \quad , \tag{4.24}$$

with $\zeta_{12}^a \equiv +2i(\bar\epsilon_2 \gamma^a \epsilon_1)$ like the component case (2.3).

Note that since we are dealing with superspace, we have an additional manifest linear supersymmetry $\delta_{Q'}$ dictated by [41]

$$\delta_{Q'} \Lambda^{\underline{\alpha}} = -\epsilon'^{\underline{\beta}} D^{(0)}_{\underline{\beta}} \Lambda^{\underline{\alpha}} \quad , \quad \delta_{Q'} \Phi = -\epsilon'^{\underline{\beta}} D^{(0)}_{\underline{\beta}} \Phi \quad , \tag{4.25}$$

with $D^{(0)}_A \equiv E^{(0)M}_A \partial_M$. The commutator of two such linear supersymmetries induces the familiar translation operator:

$$[\delta_{Q'}(\epsilon'_1), \delta_{Q'}(\epsilon'_2)] = +\delta_P(\zeta'^a_{12}) \quad , \tag{4.26}$$

where $\zeta'^a_{12} \equiv +2i(\bar\epsilon'_2 \gamma^a \epsilon'_1)$ similarly to our non-linear supersymmetry (2.3). Interestingly, by simple commutator computation on $\Lambda^{\underline{\alpha}}$ and $\Phi$, we see that these two sorts of supersymmetries are commuting each other:

$$[\delta_Q(\epsilon), \delta_{Q'}(\epsilon')] = 0 \quad , \tag{4.27}$$

as desired, with no interference.

Having seen the existence of two sorts of supersymmetries, a natural question is whether or not these two supersymmetries are equivalent to each other after all, and if not, how the usual theorem about maximal number of supersymmetries [47][48] in each space-time dimension can be avoided? Our answer to the first question is that these two supersymmetries are not equivalent to each other connected by super Weyl rescaling (field redefinitions), because the vielbeins (4.17) for the supersymmetry $\delta_Q$ gives geometry in superspace different from the 'flat' one $\delta_{Q'}$. As for the second question, we understand that the usual argument on maximal number of supersymmetries is based on the 'linear' representation that mixes physical fields with helicity differences of 1/2. Since the non-linear representation does not shuffles such helicities, the usual restriction on the maximal number of extended supersymmetries seems to be avoided here. This is the reason why we can set up two different supersymmetries at the same time, as a (Supersymmetry)$^2$-model.

We can also postulate a possible action as a superspace generalization of $I_E$ in (2.7):[21]

$$I_E \equiv \int d^D x\, d^{\widetilde{N}}\theta\, \text{sdet}\,(E_M{}^A) \equiv \int d^D x\, d^{\widetilde{N}}\theta\, E^{-1} \quad, \tag{4.28}$$

where $\widetilde{N} \equiv 2^{[D/2]} N$ is the number of the $\theta$-coordinates for $^\forall N$ supersymmetry in $^\forall D$, counting each component of a Majorana spinor as unity. However, we immediately notice that such an action will produce higher-derivative terms, as the mass dimension of the $\theta$-integral indicates, as usual in superspace [41]. In fact, the superdeterminant is expanded as

$$E^{-1} = 1 + i(\overline{\Lambda}\gamma^a D_a \Lambda) + \mathcal{O}(\Lambda^4) \quad, \tag{4.29}$$

with $D_a \equiv E_a{}^M \partial_M$. As a simple dimensional analysis reveals, some of the component fields in $\Lambda^{\underline{\alpha}}$ have higher-order derivatives in their kinetic terms.

In order to overcome this problem, we propose two models. The first one is to introduce some 'spurion' scalar superfields [49] $\Phi$ that is to be inserted into the above lagrangian:

$$I_{E\Phi} = \int d^D x\, d^{\widetilde{N}}\theta\, E^{-1} \Phi = \int d^D x\, [\, i d_0 (\overline{\lambda}\gamma^a D_a \lambda) + \mathcal{O}(\Lambda^4) \,] \quad, \tag{4.30}$$

with

$$\Phi = \theta^{\widetilde{N}} d_0 \quad, \tag{4.31}$$

where $d_0 \neq 0$ is an analog of the $D$-component in $D = 4$ for the highest sector component field of $\Phi$, and we identify $\Lambda| \equiv \lambda$. The kinetic term in (4.30) is physical with no negative energy ghosts or higher-derivatives involved. Even though this first model has the drawback due to the explicit breaking of both supersymmetries, it may still be of some mathematical interest, because of the co-existence of two sorts of supersymmetries. We expect more improved models to be developed in order to avoid such explicit breakings.

Our second model is much more advanced. We introduce a constraint action with a Lagrange multiplier superfield $L^{\underline{\alpha}M\underline{\beta}N}$ for a bilinear combination of $\Lambda^{\underline{\alpha}}$ [50]:

$$I_{L\Lambda^2} \equiv \int d^D x\, d^{\widetilde{N}}\theta\, \tfrac{1}{2} E^{-1} L^{\underline{\alpha}a\underline{\beta}b} (D_b \Lambda_{\underline{\beta}})(D_a \Lambda_{\underline{\alpha}}) \quad, \tag{4.32}$$

where $L$ has no (anti)symmetry other than $L^{\underline{\alpha}a\underline{\beta}b} = -L^{\underline{\beta}b\underline{\alpha}a}$. Accordingly, the $L$ transforms in the standard fashion:

$$\delta_Q L^{\underline{\alpha}a\underline{\beta}b} = \xi^M \partial_M L^{\underline{\alpha}a\underline{\beta}b} = \delta_{\text{G}}(\xi^M) L^{\underline{\alpha}a\underline{\beta}b} \quad,$$
$$\delta_{Q'} L^{\underline{\alpha}a\underline{\beta}b} = -\epsilon'^{\underline{\gamma}} D^{(0)}_{\underline{\gamma}} L^{\underline{\alpha}a\underline{\beta}b} \quad, \tag{4.33}$$

---

[21]To comply with the superspace notation in [41], we use $E \equiv \text{sdet}\,(E_A{}^M)$ with the inverse power compared with our component convention.

so that the total action $I \equiv I_E + I_{L\Lambda^2}$ is invariant under both $\delta_Q$ and $\delta_{Q'}$. Notice here that the derivative in $I_{L\Lambda^2}$ deletes the effect of the inhomogeneous term in the transformation of $\Lambda$ (4.19). Clearly, the superfield equation for $L$ from the total action $I$ implies the vanishing of $D_a \Lambda_{\underline{\alpha}}$:

$$\frac{\delta}{\delta L^{\underline{\alpha}a\underline{\beta}b}} I = (D_b \Lambda_{\underline{\beta}})(D_a \Lambda_{\underline{\alpha}}) = 0 \implies D_a \Lambda_{\underline{\alpha}} = 0 \ , \tag{4.34}$$

because all the indices $a, \underline{\alpha}, b, \underline{\beta}$ are free. This superfield equation preserves both the non-linear $\delta_Q$ and linear $\delta_{Q'}$ supersymmetries. In particular, the effect of the inhomogeneous term in (4.19) disappears under the derivative. On the other hand, the $\Lambda$-superfield equation is

$$\begin{aligned}\frac{\delta}{\delta \Lambda_{\underline{\alpha}}} I =\ & E^{-1}(-1)^M \Big(\frac{\delta E_M{}^{\underline{\beta}}}{\delta \Lambda_{\underline{\alpha}}}\Big) \Big[ E_{\underline{\beta}}{}^M + \tfrac{1}{2} E_{\underline{\beta}}{}^M L^{\underline{\gamma}c\underline{\delta}d}(D_d \Lambda_{\underline{\delta}})(D_c \Lambda_{\underline{\delta}}) \Big] \\ & - D_a(E^{-1} L^{\underline{\alpha}a\underline{\beta}b} D_b \Lambda_{\underline{\beta}}) + \mathcal{O}(\Lambda^5) \ . \end{aligned} \tag{4.35}$$

Since $D_a \Lambda_{\underline{\alpha}} = 0$ on-shell, each term in (4.35) vanishes, and there arises no problem of higher-derivative kinetic term for $\Lambda$. As usual in this sort of formulation, $L^{\underline{\alpha}a\underline{\beta}b}$ does not enter any superfield equations effectively, once (4.34) is satisfied. Even if we add other 'matter' actions to the above total action $I$, the NG-fermion is 'frozen' with no dynamics.

The effective disappearance of the multiplier $L^{\underline{\alpha}a\underline{\beta}b}$ from the set of physical field equations suggests the existence of appropriate gauge transformation that can eliminate $L^{\underline{\alpha}a\underline{\beta}b}$ [50]. In fact, it is easy to show that the total action $I = I_E + I_{L\Lambda^2}$ is invariant up to $\mathcal{O}(\Lambda^4)$-terms under the local $\eta$-symmetry:

$$\begin{aligned}\delta_\eta \Lambda_{\underline{\alpha}} &= \eta^b D_b \Lambda_{\underline{\alpha}} \ , \\ \delta_\eta L^{\underline{\alpha}a\underline{\beta}b} &= \big[\, 2i\eta^a (\gamma^b)^{\underline{\alpha}\underline{\beta}} - (D_c \eta^a) L^{\underline{\beta}b\underline{\alpha}c} + \tfrac{1}{2} D_c(\eta^c L^{\underline{\alpha}a\underline{\beta}b}) \big] - (\underline{\alpha}a \leftrightarrow \underline{\beta}b) + \mathcal{O}(\Lambda^2) \ , \end{aligned} \tag{4.36}$$

where $\eta^a$ is a local vectorial parameter.

This second model (4.32) with the constraint action $I_{L\Lambda^2}$ is more complicated than the first one (4.30), due to the supersymmetric invariance maintained at the action as well as the superfield equation level. Similarly to the component case in section 2 for ordinary space-time, the co-existence of two sorts of superspaces is common to these two models, one with the usual flat superspace metric, and the other with non-vanishing curvature with non-trivial vielbeins. To our knowledge, these models here are the first examples of this kind with the realization of two different supersymmetries at the same time. In particular, in the second model, both of them are exact at the action level.

We draw readers' attention to the point that our formulation with the unbroken non-linear supersymmetry is distinguished from other formulations with the non-linear realization as a manifestation of spontaneously 'broken' linear supersymmetry, such as in refs. [17][18]. The interesting aspect of our formulation is that both non-linear and linear supersymmetries coexist in superspace with no mutual inconsistency, that was not emphasized before.

## 5. Concluding Remarks

In this paper we have first studied non-linear realization of $^\forall N$-extended global supersymmetry in arbitrary dimensions $D$, and we have given actions invariant under non-linear supersymmetry transformations. These actions and transformations are valid in *arbitrary* dimensions $D$ and for *arbitrary* $N$, including of course $D \geq 12$, or $N \geq 9$ in $D = 4$. This is because the vielbein field (2.5) transforms like the general coordinate transformation with the parameter $\xi^\mu \equiv i(\bar{\epsilon}\gamma^\mu\lambda)$ under the non-linear supersymmetry, even though supersymmetry is global. The action $I_E$ has non-trivial features, because of the kinetic term of the NG-fermion, with infinitely many interaction terms, despite of its superficially simple form. Even though it seems almost straightforward to generalize the original VA action [1] to $^\forall D$ and $^\forall N$, these generalizations will be of importance for later sections.

These formulation have no need for the bosonic superpartner fields that are indispensable in other formulations, such as in [17], independent of $D$ or $N$. This feature is very similar to the original model by Volkov-Akulov [1] with the single NG-fermion. Reviewing also the transformation rule given in [9], we notice that the vector field does not enter the transformation rule of the NG-fermion, and this signals the decoupling of the vector field from the NG-fermion system. From this viewpoint, it is natural that such a vector field can be completely decoupled from the NG-fermion system, in some choice of frame without spoiling global supersymmetry.

For linear representation of supersymmetry, interacting models are possible only $D \leq 11$, unless Lorentz symmetry is broken, such as in those formulations in [46]. Therefore it has been believed that any putative interacting supersymmetric model in $D \geq 12$ with Lorentz covariance is 'doomed' to fail. The result in this paper have broken this 'taboo', namely we have seen that it is certainly possible to formulate global non-linear supersymmetry in $D \geq 12$, as long as spinor fields can be defined. In other words, global supersymmetry and Lorentz symmetry can co-exist for any given $D$ and $N$, only at the expense of the linear realization of local supersymmetry. We now know that interacting models with non-linear realization of global supersymmetries are possible even for $N \geq 9$ in $D = 4$, or $N \geq 2$ in $D = 11$, or in $D \geq 12$. Since the non-linear realization of supersymmetry is possible even in $D \geq 12$ with global Lorentz symmetry, we do not take the standpoint that non-linear supersymmetry is a reflection of 'broken' supersymmetries [17].

Although it emerged just as an interesting by-product, the supersymmetric Born-Infeld action for a non-Abelian gauge group for $^\forall D$ and $^\forall N$ we gave is a new result in this paper. Like other actions, this action also has non-trivial but consistent interactions with the NG-fermion, and it agrees in the special case of $D = 10$ with the result in [9] for the lowest-order terms. When the NG-fermion carries the adjoint index of the non-Abelian group, the lowest-order terms of our Born-Infeld action (3.6) coincide with those in the linearly supersymmetric action [9] in $D = 10$. This seems to imply that our Born-Infeld action (3.6) provides an equally important foundation for the M(atrix)-theory [27] as the linearly supersymmetric Born-Infeld action [9] for D-branes [10][11][12][14], up to some recently-discovered subtlety at higher orders [29]. Interestingly enough, since we have $\dim G$-extended supersymmetry, the usual large $\mathcal{N}$ limit for $G = U(\mathcal{N})$ [27] corresponds to $N = \mathcal{N}^2 \to \infty$, *i.e.*, infinitely many ($\aleph_0$) extended supersymmetry limit for our non-linear realization.

Additionally, we have shown how to perform duality transformation from an Abelian field strength $F_{\mu\nu}$ in a Born-Infeld (type) action into its Hodge-dual field strength $N_{\mu_1\cdots\mu_{D-2}}$ in $^\forall D \neq 4$ and $^\forall N$, excluding the case of $D = 4$, in which certain subtlety arises related to the $SL(2, \mathbb{R})$ duality [40].

As another important application, we have pointed out that our supersymmetric Born-Infeld action $I_{\text{NASBI}}$ is a suitable 'matter' action compatible with Chern-Simons actions, in which there is no fundamental metric from the outset. The gauge groups can be even a supergroup such as $OSp(N|m)$, *etc.* Another important aspect is the fact that any Chern-Simons action (3.16) for any supergroup formulated in $^\forall D = 2n+1$ [30][31][32] has hidden global supersymmetry under (3.1b) realized non-linearly. This may well provide the long-standing 'missing link' between the Chern-Simons theories of supergroup and Poincaré supergravity with diffeomorphism [38], especially in the context of the superalgebra in M-theory [31][32][37].

In our formulation, we found that the minimal field content for supersymmetric Born-Infeld action is the NG-fermion and a vector field, both for Abelian and non-Abelian cases. If we perform simple dimensional reduction from the starting dimensions $D$ into one dimension lower: $D-1$, then one component of the vector field becomes a scalar field, so that more and more scalars are generated, if we go further down [9]. However, this is not the only scenario we can think of, because in our formulation, at each dimension $D$ we can formulate a supersymmetric Born-Infeld action with the minimal set of fields, *i.e.*, one NG-fermion and a vector field, but no scalars or any other fields. This is also one of the distinctive features of our formulation compared with other D-brane formulations [9][11].

Our result of extended supersymmetries on $^\forall N$ has other analogs. For example, similar situations have been already presented for infinitely many local supersymmetries in $D = 3$ [51] or arbitrarily many global supersymmetries in $D = 1$ [52]. These so-called $\aleph_0$ aspects of supersymmetries have already shown up in many different contexts in the past, and now we have explicit models at hand that elucidate such 'universality' more explicitly in $^\forall D$.

We have also reformulated the component results in superspace, and in particular, confirmed the Bianchi identities for a second-rank superfield strength for a non-Abelian group and a third-rank superfield strength, as explicit examples. We have seen that the components $F_{\underline{\alpha\beta}}{}^I$ or $G_{\underline{\alpha\beta\gamma}}$ are needed, and the special factor $P_{\underline{\alpha}}{}^b \equiv i(\gamma^b \lambda)_{\underline{\alpha}}$ is involved in a peculiar way, as a generalization of those given in [12]. Our constraints have essentially off-shell structure, namely the NG-fermion equations are not implied by the Bianchi identities, as opposed to the usual on-shell superspace formulations in higher-dimensions, or constrained superfield formulations in $D = 4$ [2][17]. Remarkably, these constraints are valid for $^\forall D$ and $^\forall N$, which, to our knowledge, have not been emphasized in the literature.

As another important result, we have presented an universal Green-Schwarz $\sigma$-model super $(2k-1)$-brane action [21][22] in $D=2k$-dimensions that can accommodate even curved superspace backgrounds, when we need to include supergravity, by introducing the maximally antisymmetric potential superfield $C_{a_1 \cdots a_{2k}}$ [22]. The requirement of satisfaction of $T$-Bianchi identities for supergravity with Lorentz covariance restricts the dimensionality $D$ to be $D=2k$ $(1 \leq k \leq 5)$.

This $(2k-1)$-brane formulation suggests the significance of the target space-time whose dimensionality coincides with that of the world-supervolume. Motivated by this, we have given a 'simplified' (Supersymmetry)$^2$-models as another by-product of our formulations, in which linear and non-linear representations of supersymmetries coexist at the same time. This is a simplified version of more general (Supersymmetry)$^2$-models, initiated in 2D [23]. We have seen that the original ansatz of forming a vielbein in terms of NG-fermion is directly generalized into superfields, and we have also given some explicit models with the simultaneous realization of both linear and non-linear supersymmetries. In particular, the second model has both of these supersymmetries explicit at the action level. In this sense, our formulation is distinguished from others, such as refs. [17][18], in which non-linear supersymmetry is regarded as spontaneous 'breaking' of linear supersymmetry.

We saw that neither the metric nor the vielbein field in our formulation is an elementary field, but they are effectively composite in terms of the NG-fermion. From this viewpoint, our theory is similar to the Bars and MacDowell theory [53] in which the metric tensor is not an elementary field, but is a composite bilinear form of gravitino which is the fundamental field. The main difference, though, is that the formulation in [53] contains a gravitino as a gauge field for the local supersymmetry, while in our theory supersymmetry is global but is realized non-linearly. Most of our actions in general space-time dimensions are only globally supersymmetric in flat $D$ dimensions, except the super $(2k-1)$-brane actions which can accommodate curved superspace backgrounds.

One might say that a possible drawback in our formulation is the loss of 'uniqueness' of actions for particle physics. This is because there are essentially no restrictions on possible invariant lagrangian by supersymmetry, as long as the general covariance $\delta_{\rm G}$ is maintained in terms of our 'metric' or 'vielbein' for $\forall D$ and $\forall N$. However, we instead stress the importance of 'universality' of supersymmetry more than 'uniqueness' of a particle theory. Our result here opened a completely new avenue, *i.e.*, supersymmetry does not necessarily restrict the space-time dimensions for possible interacting models for physics. To our knowledge, the significance of such universality has have been overlooked, or at least not emphasized during the decades since the first work in [1].

One final point to be stressed is the co-existence of two sorts of space-times or even superspaces simultaneously, which are not equivalent to each other: One with flat metric, while the other with highly non-trivial vielbeins with general coordinate transformations. We have seen this for component as well as superspace formulations. However, this is in a sense puzzling, because it is well-known in mathematics to be impossible to embed a 'curved' space-time into a 'flat' space-time of the same dimensionality. Our solution to this puzzle is that the curved space-time is 'non-perturbatively' realized from the 'flat' space-time, by adding up infinite series in terms of NG-fermions.

Moreover, the co-existence of two different superspaces is also related to the co-existence of two sorts of supersymmetries, one in linear and the other in non-linear representations. These aspects of non-linear supersymmetry may well be different manifestations of the intricate feature of M-theory [15] for curved supergravity in $D = 11$ realized first as a global supersymmetry in flat space-time, *via* a simple matrix theory [27].

We believe that the results given in this paper can provide a first step into the studies of more general features of non-linear supersymmetries in $\forall D$ and $\forall N$, motivated also by the recent development of D-branes [10]. We have even seen that the non-linear supersymmetry can co-exist with linear supersymmetry as simple $(\text{Supersymmetry})^2$-model in superspace. We have also seen that the non-linear realization as compared with the linear realization of supersymmetry has more freedom for accommodating a much wider class of models.

We are grateful to J. Bagger, S.J. Gates, Jr., F. Gonzalez-Rey, M. Luty, M. Roček, J.H. Schwarz and W. Siegel for helpful discussions. Special acknowledgement is for S.J. Gates, Jr. for many suggestions to improve the paper.

*Note Added:* After completing this paper, we have encountered a recent work by P.C. West [54], in which superembedding [14] for non-linear realizations is performed by enlarging the automorphism group of supersymmetry.